\documentclass[a4paper,reqno,12pt]{amsart}
\usepackage[all]{xy}           
\usepackage{amssymb}           
\usepackage[utf8]{inputenc}
\usepackage[T1]{fontenc}
\usepackage{hyperref}
\usepackage{tikz-cd}
\usepackage{eucal}
\usepackage{epsfig}
\usepackage{pst-grad} 
\usepackage{pst-plot} 
\numberwithin{equation}{section}

\newtheorem{definition}{Definition}[section]

\newtheorem{theorem}[definition]{Theorem}
\newtheorem{proposition}[definition]{Proposition}
\newtheorem{corollary}[definition]{Corollary}
\newtheorem{remarkth}[definition]{Remark}
\newtheorem{example}[definition]{Example}
\newenvironment{remark}{\begin{remarkth}\upshape}{\hfill$\diamond$\end{remarkth}}

\renewcommand{\emph}[1]{{\bfseries\itshape{#1}}}

\makeatletter
\newcommand\prol{\@ifstar{\@proldf}{\@prolpf}}  
\def\@prolpf{\@ifnextchar[{\@prolpf@wrt}{\@prolpf@}}
\def\@prolpf@wrt[#1]#2{\@ifnextchar[{\@prolpf@wrt@at{#1}{#2}}{\@prolpf@wrt@{#1}{#2}}}
\def\@prolpf@wrt@at#1#2[#3]{\prolsymbol^{#1}_{#3}#2}
\def\@prolpf@wrt@#1#2{\prolsymbol^{#1}#2}
\def\@prolpf@#1{\@ifnextchar[{\@prolpf@at{#1}}{\@prolpf@@{#1}}}
\def\@prolpf@at#1[#2]{\prolsymbol_{#2}#1}
\def\@prolpf@@#1{\prolsymbol#1}
\def\@proldf{\@ifnextchar[{\@proldf@wrt}{\@proldf@}}
\def\@proldf@wrt[#1]#2{\@ifnextchar[{\@proldf@wrt@at{#1}{#2}}{\@proldf@wrt@{#1}{#2}}}
\def\@proldf@wrt@at#1#2[#3]{\prolsymbol^{*#1}_{#3}#2}
\def\@proldf@wrt@#1#2{\prolsymbol^{*#1}#2}
\def\@proldf@#1{\@ifnextchar[{\@proldf@at{#1}}{\@proldf@@{#1}}}
\def\@proldf@at#1[#2]{\prolsymbol^*_{#2}#1}
\def\@proldf@@#1{\prolsymbol^*#1}
\def\prolsymbol{\mathcal{T}}
\makeatother

\begin{document}

\title[Characteristic distribution]{Characteristic distribution: An application to material bodies}

\author[V. M. Jim\'enez]{V\'ictor Manuel Jim\'enez}
\address{V\'ictor Manuel Jim\'enez:
Instituto de Ciencias Matem\'aticas (CSIC-UAM-UC3M-UCM),
c$\backslash$ Nicol\'as Cabrera, 13-15, Campus Cantoblanco, UAM
28049 Madrid, Spain} \email{victor.jimenez@icmat.es}

\author[M. de Le\'on]{Manuel de Le\'on}
\address{Manuel de Le\'on: Instituto de Ciencias Matem\'aticas (CSIC-UAM-UC3M-UCM),
c$\backslash$ Nicol\'as Cabrera, 13-15, Campus Cantoblanco, UAM
28049 Madrid, Spain} \email{mdeleon@icmat.es}

\author[M. Epstein]{Marcelo Epstein}
\address{Marcelo Epstein:
Department of Mechanical Engineering. University of Calgary. 2500 University Drive NW, Calgary, Alberta, Canada, T2N IN4} \email{epstein@enme.ucalgary.ca}

\thanks{This work has been partially supported by MINECO Grants  MTM2016-76-072-P and the ICMAT Severo Ochoa projects SEV-2011-0087 and SEV-2015-0554.
V.M.~Jiménez wishes to thank MINECO for a FPI-PhD Position and the referee for the suggestions.}
 \subjclass[2000]{}

\begin{abstract}
Associated to each material body $\mathcal{B}$ there exists a groupoid $\Omega \left( \mathcal{B} \right)$ consisting of all the material isomorphisms connecting the points of $\mathcal{B}$. The uniformity character of $\mathcal{B}$ is reflected in the properties of $\Omega \left( \mathcal{B} \right)$: $\mathcal{B}$ is uniform if, and only if, $\Omega \left( \mathcal{B} \right)$ is transitive. Smooth uniformity corresponds to a Lie groupoid and, specifically, to a Lie subgroupoid of the groupoid $\Pi^1\left({\mathcal B}, {\mathcal B} \right)$ of 1-jets of $\mathcal B$. We consider a general situation when $\Omega \left( \mathcal{B} \right)$ is only an algebraic subgroupoid. Even in this case, we can cover $\mathcal{B}$ by a material foliation whose leaves are transitive. The same happens with $\Omega \left( \mathcal{B} \right)$ and the corresponding leaves generate transitive Lie groupoids (roughly speaking, the leaves covering $\mathcal B$). This result opens the possibility to study the homogeneity of general material bodies using geometric instruments.
\end{abstract}

\maketitle

\tableofcontents

\section{Introduction}
As it is well-known, associated to any simple material body $\mathcal{B}$ there exists a groupoid $\Omega \left( \mathcal{B} \right)$ over $\mathcal{B}$ called the \textit{material groupoid of $\mathcal{B}$} (see for example  \cite{MELZA}, \cite{MEPMDL}, \cite{MEPMDLSEG} or \cite{VMJIMM}). A material body is {\it simple} (or of grade 1) if the mechanical response functional at each point depends on the deformation gradient alone (and not on higher gradients). $\Omega \left( \mathcal{B} \right)$ consists of all linear isomorphisms $P$ between the tangent spaces $T_{X} \mathcal{B}$ and $T_{Y} \mathcal{B}$ such that
$$ W \left( F  P , X \right) = W \left( F, Y \right),$$
for any deformation gradient $F$ at $Y$, where $X,Y$ run along the body $\mathcal{B}$. Here, $W$ is the mechanical response of the body $\mathcal{B}$ tipically the stored energy per unit mass.\\
\indent{The uniformity of $\mathcal{B}$ is reflected on the properties of the material groupoid $\Omega \left( \mathcal{B} \right)$. In particular, $\mathcal{B}$ is (smoothly) uniform if, and only if, $\Omega \left( \mathcal{B} \right)$ is a transitive (Lie) subgroupoid of $\Pi^{1} \left( \mathcal{B} , \mathcal{B} \right)$, where $\Pi^{1} \left( \mathcal{B} , \mathcal{B} \right)$ is the Lie groupoid over $\mathcal{B}$, called \textit{$1-$jets groupoid on $\mathcal{B}$}, of all linear isomorphisms $P$ between the tangent spaces $T_{X} \mathcal{B}$ and $T_{Y} \mathcal{B}$, for $X,Y \in \mathcal{B}$.\\
\indent{In this paper, we consider a more general situation. We study the problem from a purely mathematical framework, since we are convinced that this analysis should be relevant not only for its applications to Continuum Mechanics, but also for the general theory of groupoids.}\\
\indent{So, let $\overline{\Gamma} \subseteq \Gamma$ be a subgroupoid of a Lie groupoid $\Gamma \rightrightarrows M$; notice that we are not assuming, in principle, any differentiable structure on $\overline{\Gamma}$. Even in that case, we can construct a generalized distribution $A \overline{\Gamma}^{T}$ over $\Gamma$ generated by the (local) left-invariant vector fields on $\Gamma$ whose flow at the identities is totally contained in $\overline{\Gamma}$. This distribution $A \overline{\Gamma}^{T}$ will be called the \textit{characteristic distribution of $\overline{\Gamma}$}. Due to the groupoid structure, we can still associate two new objects to $A \overline{\Gamma}^{T}$, denoted by $A \overline{\Gamma}$ and $A \overline{\Gamma}^{\sharp}$, and defined by the following diagram:

\begin{large}
\begin{center}
 \begin{tikzcd}[column sep=huge,row sep=huge]
\Gamma\arrow[r, "A \overline{\Gamma}^{T}"] &\mathcal{P} \left( T \Gamma \right) \arrow[d, "T\alpha"] \\
 M \arrow[u,"\epsilon"] \arrow[r,"A \overline{\Gamma}^{\sharp}"] \arrow[ru,dashrightarrow, "A \overline{\Gamma}"]&\mathcal{P} \left( T M \right)
 \end{tikzcd}
\end{center}
\end{large}
\noindent{Here $\mathcal{P} \left( E \right)$ defines the power set of $E$, $\epsilon \left( x \right)$ the unit for an element $x \in  M$ and $\alpha, \beta : \Gamma \rightrightarrows M$ denote the source and the target maps respectively. Therefore, for each $x \in M$, we have}
\begin{eqnarray*}
A \overline{\Gamma}_{x} &=&  A \overline{\Gamma}^{T}_{\epsilon \left( x \right)}\\
A \overline{\Gamma}^{\sharp} _{x}  &=& T_{\epsilon \left( x \right) } \alpha \left( A \overline{\Gamma}_{x} \right)
\end{eqnarray*}
\noindent{$A \overline{\Gamma}^{\sharp}$ is called the \textit{base-characteristic distribution of $\overline{\Gamma}$} and it is a \linebreak generalized distribution (in the sense of Stefan and Sussmann) on $M$.\\
The relevant fact is that both distributions, $A \overline{\Gamma}^{T}$ and $A\overline{\Gamma}^{\sharp}$, are integrable (in the sense of Stefan and Sussmann), and they provide two foliations, $\overline{\mathcal{F}}$ on $\Gamma$ and $\mathcal{F}$ on $M$.\\
In this paper, we have studied the properties of these foliations and obtained the following two main results:

\vspace{0.4cm}

\textbf{Theorem \ref{24}}
Let $\Gamma \rightrightarrows M$ be a Lie groupoid and $\overline{\Gamma}$ be a subgroupoid of $\Gamma$ (not necessarily a Lie groupoid) over $M$. Then, there exists a foliation $\overline{\mathcal{F}}$ of $\Gamma$ such that $\overline{\Gamma}$ is a union of leaves of $\overline{\mathcal{F}}$.

\vspace{0.4cm}

\textbf{Theorem \ref{20}}
For each $x \in M$ there exists a transitive Lie subgroupoid $\overline{\Gamma} \left( \mathcal{F} \left( x \right) \right)$ of $\Gamma$ with base $\mathcal{F} \left( x \right)$.

\vspace{0.4cm}

So, although our groupoid $\overline{\Gamma}$ is not a Lie subgroupoid of $\Gamma$, we can still cover it by manifolds (leaves of the foliation $\overline{\mathcal{F}}$) and extract ``transitive" and ``differentiable" components (the Lie groupoids $\overline{\Gamma} \left( \mathcal{F} \left( x \right) \right) \rightrightarrows \mathcal{F} \left( x \right)$).\\
\indent{The next step is to apply our result to the theory a simple bodies in Continuum Mechanics. In particular, let $\mathcal{B}$ be a simple material and $\Omega \left( \mathcal{B} \right)$ its material groupoid. Then, $\Omega \left( \mathcal{B} \right)$ is not necessarily a Lie subgroupoid of $\Pi^{1} \left( \mathcal{B} , \mathcal{B} \right)$. But, applying the results of the previous section we have that $\mathcal{B}$ can be covered by a foliation of some kind of smoothly uniform ``subbodies" (these are not exactly subbodies in the usual sense of continuum mechanics \cite{CCWAN} because of the dimension), called \textit{material submanifolds}.}\\
\indent{Finally, we present several examples in which the material groupoid is not a Lie subgroupoid of $\Pi^{1} \left( \mathcal{B} , \mathcal{B} \right)$. In each case, we give explicitily the characteristic foliation which decomposes the body into smoothly uniform material submanifolds.}\\
\indent{The paper is structured as follows: In section 2 we give a brief introduction to (Lie) groupoids (see \cite{KMG} or \cite{JNM} for a detailed account of the theory of groupoids). Section 3 is devoted to develop the theory of the characteristic distribution and prove the two main theorems. In section 4 we apply these results to continuum mechanics. Finally, some examples are discussed in Section 5.}

\section{Groupoids}

First, we shall give a brief introduction to \textit{Lie groupoids}. The standard reference on groupoid is \cite{KMG}; for a short introduction see \cite{JNM}.

\begin{definition}
\rm
Let $ M$ be a set. A \textit{groupoid} over $M$ is given by a set $\Gamma$ provided with two maps $\alpha,\beta : \Gamma \rightarrow M$ (\textit{source} and \textit{target maps}, respectively), $\epsilon: M \rightarrow \Gamma$ (\textit{identities map}), $i: \Gamma \rightarrow \Gamma$ (\textit{inversion map}) and   
$\cdot : \Gamma_{\left(2\right)} \rightarrow \Gamma$ (\textit{composition law}) where for each $k \in \mathbb{N}$,
$$\Gamma_{\left(k\right)} := \{ \left(g_{1}, \hdots , g_{k}\right) \in \Gamma^{k}  \ : \ \alpha\left(g_{i}\right)=\beta\left(g_{i+1}\right), \ i=1, \hdots , k -1 \},$$ satisfying the following properties:\\
\begin{itemize}
\item[(1)] $\alpha$ and $\beta$ are surjective and, for each $\left(g,h\right) \in \Gamma_{\left(2\right)}$, we have
$$ \alpha\left(g \cdot h \right)= \alpha\left(h\right), \ \ \ \beta\left(g \cdot h \right) = \beta\left(g\right).$$
\item[(2)] Associativity of the composition law, i.e.,
$$ g \cdot \left(h \cdot k\right) = \left(g \cdot h \right) \cdot k, \ \forall \left(g,h,k\right) \in \Gamma_{\left(3\right)}.$$
\item[(3)] For all $ g \in \Gamma$,
$$ g \cdot \epsilon \left( \alpha\left(g\right)\right) = g = \epsilon \left(\beta \left(g\right)\right)\cdot g .$$
In particular,
$$ \alpha \circ  \epsilon \circ \alpha = \alpha , \ \ \ \beta \circ \epsilon \circ \beta = \beta.$$
Since $\alpha$ and $\beta$ are surjetive we get
$$ \alpha \circ \epsilon = Id_{\Gamma}, \ \ \ \beta \circ \epsilon = Id_{\Gamma}.$$
\item[(4)] For each $g \in \Gamma$,
$$i\left(g\right) \cdot g = \epsilon \left(\alpha\left(g\right)\right) , \ \ \ g \cdot i\left(g\right) = \epsilon \left(\beta\left(g\right)\right).$$
Then,
$$ \alpha \circ i = \beta , \ \ \ \beta \circ i = \alpha.$$
\end{itemize}
These maps $\alpha, \beta, i,  \epsilon$ will be called \textit{structure maps}. In what follows, we will denote this groupoid by $ \Gamma \rightrightarrows M$.
\end{definition}
If $\Gamma$ is a groupoid over $M$, then $M$ is also denoted by $\Gamma_{\left(0\right)}$ and it is often identified with the set $\epsilon \left(M\right)$ of identity elements of $\Gamma$. $\Gamma$ is also denoted by $\Gamma_{\left(1\right)}$. The map $\left(\alpha , \beta\right) : \Gamma \rightarrow M \times M$ is called the \textit{anchor} of the groupoid.\\

\vspace{0.2cm}

Now, we define the morphisms in the category of groupoids.
\begin{definition}
\rm
If $\Gamma_{1} \rightrightarrows M_{1}$ and $\Gamma_{2} \rightrightarrows M_{2}$ are two groupoids then a morphism from $\Gamma_{1} \rightrightarrows M_{1}$ to $\Gamma_{2} \rightrightarrows M_{2}$ consists of two maps $\Phi : \Gamma_{1} \rightarrow \Gamma_{2}$ and $\phi : M_{1} \rightarrow M_{2}$ such that for any $g_{1} \in \Gamma_{1}$
\begin{equation}\label{1}
\alpha_{2} \left( \Phi \left(g_{1}\right)\right) = \phi \left(\alpha_{1} \left(g_{1} \right)\right), \ \ \ \ \ \ \ \beta_{2} \left( \Phi \left(g_{1}\right)\right) = \phi \left(\beta_{1} \left(g_{1} \right)\right),
\end{equation}
where $\alpha_{i}$ and $\beta_{i}$ are the source and the target map of $\Gamma_{i} \rightrightarrows M_{i}$ respectively, for $i=1,2$, and preserves the composition, i.e.,
$$\Phi \left( g_{1} \cdot h_{1} \right) = \Phi \left(g_{1}\right) \cdot \Phi \left(h_{1}\right), \ \forall \left(g_{1} , h_{1} \right) \in \left( \Gamma_{1}\right)_{\left(2\right)}.$$
We will denote this morphism by $\left(\Phi , \phi\right)$ or by $\Phi$ (because, using Eq. \ref{1}, $\phi$ is completely determined by $\Phi$).
\end{definition}
Observe that, as a consequence, $\Phi$ preserves the identities, i.e., denoting by $\epsilon_{i}$ the section of identities of $\Gamma_{i} \rightrightarrows M_{i}$ for $i=1,2$, we have
$$\Phi \circ  \epsilon_{1} = \epsilon_{2} \circ \phi .$$
 
Using this definition we define a \textit{subgroupoid} of a groupoid $\Gamma \rightrightarrows M$ as a groupoid $\Gamma' \rightrightarrows M'$ such that $M' \subseteq M$, $\Gamma' \subseteq \Gamma$ and the corresponding inclusion map is a morphism of groupoids.

\begin{remark}
\rm
There is a more abstract way of defining a groupoid. A groupoid is a "small" category (the class of objects and the class of morphisms are sets) in which each morphism is invertible.\\
If $ \Gamma \rightrightarrows M$ is the groupoid, then $M$ is the set of objects and $\Gamma$ is the set of morphisms. In this sense, we can think about a groupoid as a set $M$ of objects and a set $\Gamma$ of invertible maps between objects of $M$. Then, for each map $g \in \Gamma$, $\alpha \left( g \right)$ is the domain of $g$, $\beta \left( g \right)$ is the codomain $g$ and $i \left( g \right)$ is the inverse of $g$. For all $x \in M$, $\epsilon \left( x \right)$ is the identity map at $x$ and, finally, the operation $\cdot$ can be thought as the composition of maps.\\
A groupoid morphism is a functor between these categories, which is a more natural definition.
\hfill
\end{remark}
Now, we present the most basic examples of groupoids.

\begin{example}\label{2}
\rm
A group is a groupoid over a point. In fact, let $G$ be a group and $e$ the identity element of $G$. Then, $G \rightrightarrows \{e\}$ is a groupoid, where the operation of the groupoid, $\cdot$, is just the operation in $G$.
\end{example}

\begin{example}\label{3}

\rm
For any set $M$, we can consider the product space $ M \times M$. Then $M \times M$ has a groupoid structure over $M$ such that
$$ \left( x , y \right) \cdot \left( z , x \right) = \left( z , y \right),$$
for all $x,y,z \in M$. $M \times M \rightrightarrows M$ is said to be the \textit{pair groupoid of $M$}.\\
Note that, if $\Gamma \rightrightarrows M$ is an arbitrary groupoid over $M$, then the anchor $\left(\alpha , \beta\right) : \Gamma \rightarrow M \times M$ is a morphism from $\Gamma \rightrightarrows M$ to the pair groupoid of $M$.
\end{example}

Next, we introduce the notion of orbits and isotropy group.
\begin{definition}
\rm
Let $\Gamma \rightrightarrows M$ be a groupoid with $\alpha$ and $\beta$ the source map and target map, respectively. For each $x \in M$, we denote
$$\Gamma_{x}^{x} = \beta^{-1}\left(x\right) \cap \alpha^{-1}\left(x\right),$$
which is called the \textit{isotropy group of} $\Gamma$ at $x$. The set
$$\mathcal{O}\left(x\right) = \beta\left(\alpha^{-1}\left(x\right)\right) = \alpha\left(\beta^{-1}\left( x\right)\right),$$
is called the \textit{orbit} of $x$, or \textit{the orbit} of $\Gamma$ through $x$.\\
\indent{If $\mathcal{O}\left(x\right) = M$ for all $x \in M$, or equivalently $\left(\alpha,\beta\right) : \Gamma  \rightarrow M \times M$ is a surjective map, then the groupoid $\Gamma \rightrightarrows M$ is called \textit{transitive}.}\\
\indent{Furthermore, the preimages of the source map $\alpha$ of a groupoid are called $\alpha-$\textit{fibres}. Those of the target map $\beta$ are called $\beta-$\textit{fibres}. We will usually denote the $\alpha-$fibre (resp. $\beta-$fibre) at a point $x$ by $\Gamma_{x}$ (resp. $\Gamma^{x}$).}
\end{definition}

\begin{definition}\label{4}
\rm  
Let $\Gamma \rightrightarrows M$ be a groupoid with $\alpha$ and $\beta$ the source and target map, respectively. We may define the left translation on $g \in \Gamma$ as the map $L_{g} : \beta^{-1} \left( \alpha\left(g\right)\right) \rightarrow \beta^{-1} \left(\beta\left(g\right)\right)$, given by
$$ h \mapsto g \cdot h .$$
Similarly, we may define the right translation on $g$, $R_{g} : \alpha^{-1}\left(\beta\left(g\right)\right) \rightarrow \alpha^{-1} \left( \alpha \left(g\right)\right)$. 
\end{definition}
Note that,
\begin{equation}\label{5} 
Id_{\beta^{-1}\left(x\right)} = L_{\epsilon \left(x\right)}.
\end{equation}
So, for all $ g \in \Gamma $, the left (resp. right) translation on $g$, $L_{g}$ (resp. $R_{g}$), is a bijective map with inverse $L_{i\left(g\right)}$ (resp. $R_{i\left(g\right)}$), where $i : \Gamma \rightarrow \Gamma$ is the inverse map.\\
\indent{Different structures (topological and geometrical) can be imposed on groupoids, depending on the context we are dealing with. We are interested in a particular case, the so-called Lie groupoids.}
\begin{definition}
\rm
A \textit{Lie groupoid} is a groupoid $\Gamma \rightrightarrows M$ such that $\Gamma$ and $M$ are smooth manifolds, and all the structure maps are smooth. Furthermore, the source and the target maps are submersions.\\
A \textit{Lie groupoid morphism} is a groupoid morphism which is differentiable.\\
\end{definition}

\begin{definition}
\rm
Let $\Gamma \rightrightarrows M$ be a Lie groupoid. A \textit{Lie subgroupoid} of $\Gamma \rightrightarrows M$ is a Lie groupoid $\Gamma' \rightrightarrows M'$ such that $\Gamma' $ and $M'$ are submanifolds of $\Gamma$ and $M$, respectively; and the pair given by the inclusion maps $j_{\Gamma'} : \Gamma' \hookrightarrow \Gamma$ $j_{M'} : M' \hookrightarrow M$ become a morphism of Lie groupoids.
\end{definition}
Observe that, taking into account that $ \alpha \circ \epsilon = Id_{M} = \beta \circ \epsilon$, then $\epsilon$ is an injective immersion.\\
\indent{On the other hand, in the case of a Lie groupoid, $L_{g}$ (resp. $R_{g}$) is clearly a diffeomorphism for every $g \in \Gamma$.}\\

\begin{example}\label{6}
\rm
A Lie group is a Lie groupoid over a point. 
\end{example}

\begin{example}
\rm
Let $M$ be a manifold. The pair groupoid $M \times M \rightrightarrows M$ is a Lie groupoid.
\end{example}
Next, we will introduce an example which will be fundamental in this paper.
\begin{example}\label{7}
\rm
Let $M$ be a manifold, and denote by $\Pi^{1} \left(M,M\right)$ the set of all vector space isomorphisms $L_{x,y}: T_{x}M \rightarrow T_{y}M$ for $x,y \in M$ or, equivalently, the space of the $1-$jets of local diffeomorphisms on $M$. An element of $\Pi^{1} \left(M,M\right)$ will by denoted by $j^{1}_{x,y}\psi$, where $\psi$ is a local diffeomorphism from $M$ into $M$ such that $\psi \left( x \right) = y$.\\
\indent{$\Pi^{1} \left(M,M\right)$ can be seen as a groupoid over $M$ where, for all $x,y \in M$ and $j^{1}_{x,y}\psi , j^{1}_{y,z}\varphi \in \Pi^{1} \left(M,M\right)$, we have}
\begin{itemize}
\item[(i)] $\alpha\left(j^{1}_{x,y}\psi\right) = x$
\item[(ii)] $\beta\left(j^{1}_{x,y}\psi\right) = y$
\item[(iii)] $j^{1}_{y,z}\varphi \cdot j^{1}_{x,y}\psi = j^{1}_{x,z} \left( \varphi \circ \psi \right)$
\end{itemize}
This groupoid is called the $1-$\textit{jets groupoid on $M$}. In fact, let $\left(x^{i}\right)$ and $\left(y^{j}\right)$ be local coordinate systems on open sets $U, V \subseteq M$. Then, we can consider a local coordinate system on $\Pi^{1} \left(M,M\right)$ given by
\begin{equation}\label{8}
\Pi^{1}\left(U,V\right) : \left(x^{i} , y^{j}, y^{j}_{i}\right),
\end{equation}
where, for each $ j^{1}_{x,y} \psi \in \Pi^{1}\left(U,V\right)$
\begin{itemize}
\item $x^{i} \left(j^{1}_{x,y} \psi\right) = x^{i} \left(x\right)$.
\item $y^{j} \left(j^{1}_{x,y}\psi \right) = y^{j} \left( y\right)$.
\item $y^{j}_{i}\left( j^{1}_{x,y}\psi\right)  = \dfrac{\partial \left(y^{j}\circ \psi\right)}{\partial x^{i}_{| x} }$.
\end{itemize}
These local coordinates turn this groupoid into a Lie groupoid.
\end{example}

\section{Characteristic distribution}
Sometimes it could be necessary to work with a groupoid which does not have a structure of Lie groupoid. In fact, the constitutive theory of continuum mechanics is an example. In this case, the set of material isomorphisms is not necessarily a Lie groupoid but it is contained in a Lie groupoid (the $1-$jets groupoid on a manifold). This will be discussed in the next section in some detail.\\
Let $ \Gamma \rightrightarrows M$ be a Lie groupoid and $\overline{\Gamma}$ be a subgroupoid of $\Gamma$ (not necessarily a Lie subgroupoid of $\Gamma$) over the same manifold $M$. We will denote by $\overline{\alpha}$, $\overline{\beta}$, $\overline{\epsilon}$ and $\overline{i}$ the restrictions of the structure maps of $\Gamma$ to $\overline{\Gamma}$ (see the diagram below).\\

\begin{center}
 \begin{tikzcd}[column sep=huge,row sep=huge]
\overline{\Gamma}\arrow[r, hook, "j"] \arrow[rd, shift right=0.5ex] \arrow[rd, shift left=0.5ex]&\Gamma \arrow[d, shift right=0.5ex] \arrow[d, shift left=0.5ex] \\
& M 
 \end{tikzcd}
\end{center}

\noindent{where $j$ is the inclusion map. Now, we can construct a distribution $A \overline{\Gamma}^{T}$ over the manifold $\Gamma$ in the following way,
$$ g \in \Gamma \mapsto A \overline{\Gamma}^{T}_{g} \leq T_{g} \Gamma,$$
such that $A \overline{\Gamma}^{T}_{g}$ is generated by the (local) left-invariant vector fields $X \in \frak X_{loc} \left( \Gamma \right)$ whose flow at the identities is totally contained in $\overline{\Gamma}$, i.e.,
\begin{itemize}
\item[(i)] $X$ is tangent to the $\beta-$fibres, 
$$ X \left( g \right) \in T_{g} \beta^{-1} \left( \beta \left( g \right) \right),$$
for all $g$ in the domain of $X$.
\item[(ii)] $X$ is invariant by left translations,
$$ X \left( g \right) = T_{\epsilon \left( \alpha \left( g \right) \right) } L_{g} \left( X \left( \epsilon \left( \alpha \left( g \right) \right) \right) \right),$$
for all $g $ in the domain of $X$.
\item[(iii)] The (local) flow $\varphi^{X}_{t}$ of $X$ satisfies
$$\varphi^{X}_{t} \left( \epsilon \left( x \right)\right) \in \overline{\Gamma}, $$
for all $x \in M$.
\end{itemize}
Notice that, for each $g \in \Gamma$, the zero vector $0_{g} \in T_{g} \Gamma$ is contained in the fibre of the distribution at $g$, namely $A \overline{\Gamma}^{T}_{g}$ (we remit to the last section for non-trivial examples). On the other hand, it is easy to prove that a vector field $X$ satisfies conditions (i) and (ii) if, and only if, its local flow $\varphi^{X}_{t}$ is left-invariant or, equivalently,
$$ L_{g} \circ \varphi^{X}_{t} = \varphi^{X}_{t} \circ L_{g}, \ \forall g,t.$$
Then, taking into account that all the identities are in $\overline{\Gamma}$ (because it is a subgroupoid of $\Gamma$), condition (iii) is equivalent to the following,
\begin{itemize}
\item[(iii)'] The (local) flow $\varphi^{X}_{t}$ of $X$ at $\overline{g}$ is totally contained in $\overline{\Gamma}$, for all $\overline{g} \in \overline{\Gamma}$.
\end{itemize}
Thus, we are taking the left-invariant vector fields on $\Gamma$ whose integral curves are confined inside or outside $\overline{\Gamma}$. It is also remarkable that, by definition, this distribution is differentiable.\\
\indent{The distribution $A \overline{\Gamma}^{T}$ is called the \textit{characteristic distribution of $\overline{\Gamma}$}.}\\
For the sake of simplicity, we will denote the family of the vector fields which satisfy conditions (i), (ii) and (iii) by $\mathcal{C}$.\\

We can still construct two new objects associated to the distribution $A \overline{\Gamma}^{T}$. The first one is a smooth distribution over the base $M$ denoted by $A \overline{\Gamma}^{\sharp}$. The second one is a ``differentiable" correspondence $A\overline{\Gamma}$ which associates to any point $x$ of $M$ a vector subspace of $T_{\epsilon \left( x \right) } \Gamma$. Both constructions are characterized by the commutativity of the following diagram

\begin{large}
\begin{center}
 \begin{tikzcd}[column sep=huge,row sep=huge]
\Gamma\arrow[r, "A \overline{\Gamma}^{T}"] &\mathcal{P} \left( T \Gamma \right) \arrow[d, "T\alpha"] \\
 M \arrow[u,"\epsilon"] \arrow[r,"A \overline{\Gamma}^{\sharp}"] \arrow[ru,dashrightarrow, "A \overline{\Gamma}"]&\mathcal{P} \left( T M \right)
 \end{tikzcd}
\end{center}
\end{large}

%

\vspace{35pt}
\noindent{where $\mathcal{P} \left( E \right)$ defines the power set of $E$. Therefore, for each $x \in M$, we have}
\begin{eqnarray*}
A \overline{\Gamma}_{x} &=&  A \overline{\Gamma}^{T}_{\epsilon \left( x \right)}\\
A \overline{\Gamma}^{\sharp} _{x}  &=& T_{\epsilon \left( x \right) } \alpha \left( A \overline{\Gamma}_{x} \right)
\end{eqnarray*}

The distribution $A \overline{\Gamma}^{\sharp}$ is called \textit{base-characteristic distribution of $\overline{\Gamma}$}. Notice that both distributions are characterized by the differentiable correspondence $A \overline{\Gamma}$ in the following way
$$ A \overline{\Gamma}^{T}_{g} = T_{\epsilon \left( \alpha \left( g \right) \right)} L_{g} \left( A \overline{\Gamma}_{ \alpha \left( g \right) } \right).$$
This equality can be proved using the construction of the associated distribution $A \overline{\Gamma}^{T}$ by left-invariant vector fields whose (local) flows at the identities are contained in $\overline{\Gamma}$. We could have used Grassmannian manifolds instead of power sets in the above diagram for the distributions but we prefer power sets because of the simplicity.\\\\
\indent{To summarize, associated to $\overline{\Gamma}$ we have three differentiable objects $A \overline{\Gamma}$, $A \overline{\Gamma}^{T}$ and $A \overline{\Gamma}^{\sharp}$. Now, we will study how these objects endow $\overline{\Gamma}$ with a sort of ``differentiable" structure.}\\

Consider a left-invariant vector field $X$ on $\Gamma$ whose (local) flow $\varphi_{t}^{X}$ at the identities is contained in $\overline{\Gamma}$. We want to prove that the characteristic distribution $A \overline{\Gamma}^T$ is invariant by the flow $\varphi_{t}^{X}$, i.e., for all $g \in \Gamma$ and $t$ in the domain of $\varphi_{g}^{X}$ we have
\begin{equation}\label{9}
T_{g} \varphi_{t}^{X} \left( A \overline{\Gamma}^{T}_{g} \right) = A \overline{\Gamma}^{T}_{\varphi_{t}^{X} \left( g \right)}.
\end{equation}

Indeed, let be $v_{g} = Y \left(g \right) \in A\overline{\Gamma}^{T}_{g}$ with $ Y \in \mathcal{C}$. Then,
\begin{eqnarray*}
T_{g} \varphi_{t}^{X} \left( v_{g} \right) &=& T_{g} \varphi_{t}^{X} \left( Y \left(g\right) \right)\\
&=&  \dfrac{\partial}{\partial s_{|0}} \left( \varphi_{t}^{X} \circ \varphi_{s}^{Y}\left( g \right) \right) ,
\end{eqnarray*}
where $\varphi^{Y}_{s}$ is the flow of $Y$.\\
Let us consider the (local) vector field on $\Gamma$ given by
$$Z \left( h \right) = \{\left( \varphi_{t}^{X} \right)^{*} Y \}\left( h \right) = T_{ \varphi^{X}_{-t}\left(h\right) } \varphi^{X}_{t} \left( Y \left( \varphi^{X}_{-t}\left(h\right) \right) \right).$$
Obviously, $Z \in \mathcal{C}$ (the flow of $Z$ is given by $\varphi_{t}^{X} \circ \varphi_{s}^{Y} \circ \varphi^{X}_{-t}$). Furthermore,
$$ T_{g} \varphi_{t}^{X} \left( v_{g} \right) = Z \left( \varphi_{t}^{X} \left( g \right) \right).$$
So, $T_{g} \varphi_{t}^{X} \left( A \overline{\Gamma}^{T}_{g} \right) \subseteq A \overline{\Gamma}^{T}_{\varphi_{t}^{X} \left( g \right)}$. We can prove the converse in an analogous way.\\

Thus, the associated distribution $A \overline{\Gamma}^{T}$ is generated by a family of vector fields $\mathcal{C}$, and it is invariant with respect this family. Now, we refer to a classical theorem, due to Stefan \cite{PS} and Sussmann \cite{HJS}, which gives an answer to the following question: what are the conditions for a smooth singular distribution to be the tangent distribution of a singular foliation?\\
\begin{theorem}[Stefan-Sussmann]\label{10}
Let $D$ be a smooth singular distribution on a smooth manifold $M$. Then the following three conditions are equivalent:
\begin{itemize}
\item[(a)] $D$ is integrable.
\item[(b)] $D$ is generated by a family $C$ of smooth vector fields, and is invariant with respect to $C$.
\item[(c)] $D$ is the tangent distribution $D^{\mathcal{F}}$ of a smooth singular foliation $\mathcal{F}$.
\end{itemize}
\end{theorem}
There is still another theorem to deal with the integrability of generalized distributions which could be confused with the Stefan-Sussmann theorem, the \textit{Hermann theorem}, that states that any locally finitely generated differentiable involutive distribution on a manifold is integrable. We refer to \cite{DUZUN} for a straightforward and clear exposition of these two theorems.\\
So, the distribution $A \overline{\Gamma}^{T}$ is the tangent distribution of a smooth singular foliation $\overline{\mathcal{F}}$. The leaf at a point $g \in \Gamma$ is denoted by $\overline{\mathcal{F}} \left( g \right)$. The collection of the leaves of $\overline{\mathcal{F}}$ which are contained in $\overline{\Gamma}$ is called the \textit{characteristic foliation of $\overline{\Gamma}$}. Note that the leaves of the characteristic foliation covers $\overline{\Gamma}$ but it is not exactly a foliation of $\overline{\Gamma}$ (because $\overline{\Gamma}$ is not manifold).\\
The following assertions can be easily proved:
\begin{itemize}
\item[(i)] For each $g \in \Gamma$,
$$\overline{\mathcal{F}} \left( g \right) \subseteq \Gamma^{\beta \left( g \right)}.$$
Indeed, if $g \in \overline{\Gamma}$, then
$$\overline{\mathcal{F}} \left( g \right) \subseteq \overline{\Gamma}^{\beta \left( g \right)}.$$
\item[(ii)] For each $g ,h \in \Gamma$ such that $\alpha \left( g \right) = \beta \left( h \right)$, we have
$$\overline{\mathcal{F}} \left( g \cdot h\right) = g \cdot \overline{\mathcal{F}} \left(  h\right).$$
\end{itemize}
The property $\left( i i \right)$ is proved by arguments of maximility. On the other hand, the property $\left( i \right)$ can be proved by checking the charts of the leaves given in the proof of the Stefan-Sussmann's theorem (see for instance \cite{MC}). It is remarkable that property $\left( i \right)$ means that each leaf of the foliation $\overline{\mathcal{F}}$ which integrates $A \overline{\Gamma}^{T}$ is contained in just one $\beta-$fibre, i.e., for each $g \in \Gamma$ the leaf $\overline{\mathcal{F}} \left( g \right)$ satisfies that 
$$\beta \left( h \right) = \beta \left( g \right),$$
for all $h \in \overline{\mathcal{F}} \left( g \right)$. Notice also that, one could expect that $\overline{\mathcal{F}} \left( g \right) = \overline{\Gamma}^{\beta \left( g \right)}$ but this is not true in general (see the example in the last section).\\

So, we have proved the following result.
\begin{theorem}\label{24}
Let $\Gamma \rightrightarrows M$ be a Lie groupoid and $\overline{\Gamma}$ be a subgroupoid of $\Gamma$ (not necessarily a Lie groupoid) over $M$. Then, there exists a foliation $\overline{\mathcal{F}}$ of $\Gamma$ such that $\overline{\Gamma}$ is a union of leaves of $\overline{\mathcal{F}}$.
\end{theorem}
In this way, without assuming that $\overline{\Gamma}$ is a manifold, we prove that $\overline{\Gamma}$ is union of leaves of a foliation of $\Gamma$. This gives us some kind of ``differentiable" structure over $\overline{\Gamma}$.\\
Let us consider a (local) left-invariant vector field $X \in \mathcal{C}$. Then, the flow of $X$ restrics to the fibers, i.e., $X$ is a left-invariant vector field in $\Gamma$ such that,
\begin{equation}\label{18}
 X_{|\overline{\mathcal{F}}\left( g \right)} \in \frak X \left( \overline{\mathcal{F}} \left( g \right) \right),
\end{equation}
for all $g $ in the domain of $X$. Reciprocally, Eq. (\ref{18}) is equivalent to conditions $(i)$, $(ii)$ and $(iii)$ which characterize the set $\mathcal{C}$.\\
An obvious consequence of the construction of the characteristic foliation is the following: \textit{ $\overline{\beta}^{-1} \left( x \right)$ is a submanifold of $\Gamma$ for all $x \in M$ if, and only if, $\overline{\beta}^{-1} \left( x \right) = \overline{\mathcal{F}} \left( \epsilon \left( x \right) \right)$ for all $x \in M$.}

\begin{remark}\label{26}
\rm
We can construct another distribution $\mathcal{D}$ on $\overline{\Gamma}$ generated by the (local) vector fields whose flows are confined inside or outside $\overline{\Gamma}$. So, we will obtain a foliation $\overline{\mathcal{G}}$ of $\Gamma$ such that $\overline{\Gamma}$ is covered by some of the leaves.\\
We could expect that the leaves at the identities $\overline{\mathcal{G}} \left( \epsilon \left( x \right) \right)$ are subgroupoids of $\Gamma$. However, this is not necessarily true. Because of this fact, we work with $A \overline{\Gamma}^{T}$ instead of $\mathcal{D}$ (see Theorem \ref{20}). 
\hfill

\end{remark}

Notice that, in an analogous way, we can prove that the base distribution $A \overline{\Gamma}^{\sharp}$ is also integrable. So, we will denote the foliation given by this distribution over the base $M$ by $\mathcal{F}$. For each point $x \in M$, the leaf of $\mathcal{F}$ through $x$ is denoted by $\mathcal{F} \left( x \right)$. $\mathcal{F}$ is called the \textit{base-characteristic foliation of $\overline{\Gamma}$}.\\\\
Next, we will prove that the leaves of $\mathcal{F}$ have even more geometric structure. In fact, we will find a Lie groupoid structure over each leaf of $\mathcal{F}$. To do this, we will prove the following technical proposition.

\begin{proposition}\label{22}
Let $\Gamma \rightrightarrows M$ be a Lie groupoid and $\overline{\Gamma}$ be a subgroupoid of $\Gamma$ with $\overline{\mathcal{F}}$ and $\mathcal{F}$ the characteristic foliation and the base-characteristic foliation of $\overline{\Gamma}$, respectively. Then, for all $x \in M$, the mapping
$$\alpha_{| \overline{\mathcal{F}} \left( \epsilon \left( x \right) \right)} : \overline{\mathcal{F}} \left( \epsilon \left( x \right) \right) \rightarrow \mathcal{F} \left( x \right),$$
is a surjective submersion.
\begin{proof}
First, let us notice that
$$ x \in \alpha \left( \overline{\mathcal{F}} \left( \epsilon \left( x \right)  \right) \right) \cap \mathcal{F} \left( x \right) \neq \emptyset.$$
Next, consider a family $\{ X^{i}, Y^{j} \}_{i= 1, \dots , r, j= 1, \dots , s}$ of left-invariant vector fields in $\mathcal{C}$ such that $\{ T_{\epsilon \left( x\right)} \alpha \left(X^{i} \left( \epsilon \left( x\right) \right) \right)\}_{i= 1 , \dots , r}$ is a basis of $A\overline{\Gamma}^{\sharp}_{ x}$ and $\{ X^{i} \left( \epsilon \left( x\right) \right) , Y^{j} \left( \epsilon \left( x\right) \right)  \}_{i= 1, \dots , r, j= 1, \dots , s}$ is a basis of $A\overline{\Gamma}^{T}_{\epsilon \left( x\right)}$.\\
Notice that the family $\{  T \alpha \circ X^{i} \circ \epsilon ,  T \alpha \circ Y^{j} \circ  \epsilon \}_{i= 1, \dots , r, j= 1, \dots , s}$ of vector fields on $M$ is tangent to the base-characteristic distribution $A \overline{\Gamma}^{\sharp}$. So, their flows at $x$ are contained in $\mathcal{F} \left( x \right)$.\\
Furthermore, the map 
$$\alpha \circ \varphi^{X^{1}}_{t_{1}} \circ \epsilon \circ \cdots \circ \alpha \circ \varphi^{X^{r}}_{t_{r}} \left( \epsilon \left( x\right) \right) = \alpha \left( \varphi^{X^{1}}_{t_{1}} \circ \cdots \circ \varphi^{X^{r}}_{t_{r}} \left( \epsilon \left( x\right) \right)\right),$$ 
defines a local chart of $\mathcal{F} \left( x \right) $ containing $x$, where $ \varphi^{X^{i}}_{t_{i}}$ is the (local) flow of $X^{i}$ for each $i$. Following this argument, one can prove that $\alpha \left( \overline{\mathcal{F}} \left( \epsilon \left( x \right)  \right) \right)$ is an open subset of $\mathcal{F} \left( x \right)$.\\
Then, $\mathcal{F} \left( x \right)$ is the disjoint union of open subsets. Using that $\mathcal{F} \left( x \right)$ is connected we have that
$$ \alpha \left( \overline{\mathcal{F}} \left( \epsilon \left( x \right)  \right) \right) = \mathcal{F} \left( x \right),$$
i.e., $\alpha_{| \overline{\mathcal{F}} \left( \epsilon \left( x \right) \right)}$ is surjective. Hence, $\alpha_{| \overline{\mathcal{F}} \left( \epsilon \left( x \right) \right)}$ is a submersion.\\
\hfill
\end{proof}
\end{proposition}
Let be $x \in M$ and $X \in \frak X \left( \mathcal{F} \left( x \right)\right)$. Then, by using local sections of $\alpha_{|\overline{\mathcal{F}} \left( \epsilon \left( x \right) \right)}$, we can extend (locally) $X$ to a vector field on $ \overline{\mathcal{F}} \left( \epsilon \left( x \right)\right)$. In this way, $X$ is a local vector field tangent to the base-characteristic distribution if, and only if, it satisfies that
\begin{equation}\label{19}
 X_{| \mathcal{F}\left(x \right)} \in \frak X \left( \mathcal{F} \left( x \right) \right),
\end{equation}
for all $x $ in the domain of $X$.\\
As a corollary, we have the following interesting result.

\begin{corollary}
Let $\Gamma \rightrightarrows M$ be a Lie groupoid and $\overline{\Gamma}$ be a subgroupoid of $\Gamma$. Then, the manifolds $\overline{\mathcal{F}} \left( \epsilon \left( x \right) \right) \cap \alpha^{-1} \left( x \right)$ are Lie subgroups of $\Gamma_{x}^{x}$ for all $x \in M$.
\begin{proof}
Let be $h, g \in \overline{\mathcal{F}} \left( \epsilon \left( x \right) \right) \cap \alpha^{-1} \left( x \right)$. Then,
$$  \overline{\mathcal{F}} \left(h \cdot g \right) = h \cdot \overline{\mathcal{F}} \left( g \right) = h \cdot \overline{\mathcal{F}} \left( \epsilon \left( x \right) \right) =  \overline{\mathcal{F}} \left( h \right) =  \overline{\mathcal{F}} \left( \epsilon \left( x \right) \right).$$
\hfill
\end{proof}

\end{corollary}

Let us denote the groupoid generated by $\overline{\mathcal{F}} \left( \epsilon \left( x \right) \right) $ by $\overline{\Gamma} \left( \mathcal{F} \left( x \right) \right)$, indeed, we construct $\overline{\Gamma} \left( \mathcal{F} \left( x \right) \right)$ in several steps:
\begin{itemize}
\item[(i)] For all $g \in \overline{\mathcal{F}} \left( \epsilon \left( x \right) \right) $,
$$ g^{-1} \in \overline{\Gamma} \left( \mathcal{F} \left( x \right) \right).$$
\item[(ii)] For all $g ,h\in \overline{\mathcal{F}} \left( \epsilon \left( x \right) \right) $, we have that:\\ 
\begin{itemize}
\item[] If $\alpha \left( h \right) = \beta \left( g \right)$,
$$h \cdot g \in \overline{\Gamma} \left( \mathcal{F} \left( x \right) \right).$$
\item[] If $\alpha \left( h \right) = \alpha \left( g \right)$,
$$h \cdot g^{-1} \in \overline{\Gamma} \left( \mathcal{F} \left( x \right) \right).$$
\item[] If $\beta \left( h \right) = \beta \left( g \right)$,
$$h^{-1} \cdot g \in \overline{\Gamma} \left( \mathcal{F} \left( x \right) \right).$$
\end{itemize}
\end{itemize}
We could define $\overline{\Gamma} \left( \mathcal{F} \left( x \right) \right)$ as the smallest subgroupoid of $\overline{\Gamma}$ which contains $\overline{\mathcal{F}} \left( \epsilon \left( x \right) \right)$. Actually, we will prove that
$$ \overline{\Gamma} \left( \mathcal{F} \left( x \right) \right) = \sqcup_{g \in \overline{\mathcal{F}} \left( \epsilon \left( x \right) \right)} \overline{\mathcal{F}} \left( \epsilon \left( \alpha \left( g \right) \right) \right),$$
or, equivalently, $\overline{\Gamma} \left( \mathcal{F} \left( x \right) \right)$ is a union of fibres at the identities (see Eq. \ref{17}).\\
\indent{This groupoid is obviously a transitive groupoid over $\mathcal{F} \left( x \right) $ and it satisfies the following properties:}
\begin{itemize}
\item[(a)] For all $g \in \overline{\mathcal{F}} \left( \epsilon \left( x \right) \right)$,
$$ \overline{\mathcal{F}} \left( g^{-1} \right) = \overline{\mathcal{F}} \left(  \epsilon \left( \alpha \left(g \right) \right) \right) \subseteq \overline{\Gamma} \left( \mathcal{F} \left( x \right) \right).$$
\item[(b)] For all $h,g \in \overline{\mathcal{F}} \left( \epsilon \left( x \right) \right)$,
$$ \overline{\mathcal{F}}\left( h^{-1} \cdot g \right)  = h^{-1} \cdot \overline{\mathcal{F}} \left( \epsilon \left( x \right) \right) \subseteq \overline{\Gamma} \left( \mathcal{F} \left( x \right) \right).$$
\item[(c)] For all $h,g \in \overline{\mathcal{F}} \left( \epsilon \left( x \right) \right)$ with $\alpha \left( g \right) = \alpha \left( h \right)$,
$$ \overline{\mathcal{F}}\left( h\cdot g^{-1} \right)  = h \cdot \overline{\mathcal{F}} \left( \epsilon \left( \alpha \left( g \right) \right) \right) \subseteq \overline{\Gamma} \left( \mathcal{F} \left( x \right) \right).$$
\item[(d)] For all $g \in \overline{\Gamma} \left( \mathcal{F} \left( x \right) \right)$,
$$ \alpha \left( \overline{\mathcal{F}} \left( g \right) \right) = \mathcal{F} \left( x \right).$$
\end{itemize}

$\left(b\right)$, $\left(c\right)$ and $\left( d \right)$ are direct consequences of $\left(a\right)$. Furthermore, for each $g \in \overline{\mathcal{F}} \left( \epsilon \left( x \right) \right)$, we have
$$ \overline{\mathcal{F}} \left( \epsilon \left( x \right) \right) = \overline{\mathcal{F}} \left( g \right) = g \cdot \overline{\mathcal{F}} \left( \epsilon \left( \alpha \left( g \right) \right) \right).$$
Then,
$$\overline{\mathcal{F}} \left( g^{-1}\right) = g^{-1} \cdot \overline{\mathcal{F}} \left( \epsilon \left( x \right)\right) =\overline{\mathcal{F}} \left( \epsilon \left( \alpha \left( g \right) \right) \right) \subseteq \overline{\Gamma} \left( \mathcal{F} \left( x \right) \right).$$
This proves $\left( a \right)$.\\
Notice that, putting together all these conditions we have that
\begin{equation}\label{17}
 \overline{\Gamma} \left( \mathcal{F} \left( x \right) \right) = \sqcup_{g \in \overline{\mathcal{F}} \left( \epsilon \left( x \right) \right)} \overline{\mathcal{F}} \left( \epsilon \left( \alpha \left( g \right) \right) \right).
\end{equation}
Futhermore, the $\beta-$fibre of this groupoid at a point $y \in \mathcal{F} \left( x \right)$ is given by $\overline{\mathcal{F}} \left( \epsilon \left( y \right) \right)$. Hence, the $\alpha-$fibre at $y$ is 
$$ \overline{\mathcal{F}}^{-1} \left( \epsilon \left( y\right) \right) = i \circ \overline{\mathcal{F}} \left( \epsilon \left( y \right) \right).$$
Thus, the fact of that the $\beta-$fibres (resp. $\alpha-$fibres) are manifolds inspires the following result. Notice also that the Lie groups $\overline{\mathcal{F}} \left( \epsilon \left( y \right) \right) \cap \alpha^{-1} \left( y \right)$ are exactly the isotropy groups of $\overline{\Gamma}\left( \mathcal{F} \left( x \right) \right) $.\\
\begin{theorem}\label{20}
For each $x \in M$ there exists a transitive Lie subgroupoid $\overline{\Gamma} \left( \mathcal{F} \left( x \right) \right)$ of $\Gamma$ with base $\mathcal{F} \left( x \right)$.
\begin{proof}
Let be $g \in \overline{\Gamma} \left( \mathcal{F} \left( x \right) \right)$. Then, by Proposition \ref{22}, the restriction
\begin{equation}\label{23}
\beta_{|\overline{\mathcal{F}}^{-1} \left( g \right)}: \overline{\mathcal{F}}^{-1} \left( g^{-1} \right) \rightarrow \mathcal{F} \left( x \right),
\end{equation}
is a surjective submersion, where $\overline{\mathcal{F}}^{-1} \left( g^{-1} \right) = i \circ \overline{\mathcal{F}} \left( g^{-1} \right)$. Using this fact, we will endow with a differentiable structure to $\overline{\Gamma} \left( \mathcal{F} \left( x \right) \right) $. Let be $g \in \overline{\Gamma} \left( \mathcal{F} \left( x \right) \right)$. Consider $\sigma_{g} : U \rightarrow \overline{\mathcal{F}}^{-1} \left( g^{-1} \right)$ a (local) section of $\beta_{|\overline{\mathcal{F}}^{-1} \left( g^{-1} \right)}$ such that $\sigma_{g} \left( \beta \left( g \right) \right) = g$.\\
On the other hand, let $\{X_{i} \}_{i=1}^{r}$ be a finite collection of vector fields in $\mathcal{C}$ such that $\{ X_{i} \left( \epsilon \left( \alpha \left( g \right) \right) \right) \}_{i=1}^{r}$ is a basis of $A \overline{\Gamma}_{\epsilon \left( \alpha \left( g \right) \right) }^{T}$. Then, a local chart over $g$ can be given by
$$
\begin{array}{rccl}
\varphi^{X} : & W \times U & \rightarrow & \Gamma \\
& \left( t_{1} , \dots, t_{r} , z \right) &\mapsto & \sigma_{g} \left( z \right) \cdot [ \varphi^{X^{r}}_{t_{r}} \circ \cdots \circ \varphi^{X^{1}}_{t_{1}} \left( \epsilon \left( \alpha \left( g \right) \right) \right)]
\end{array}
,$$
where $\varphi^{X^{i}}_{t}$ is the flow of $X^{i}$. By using that $\{  X^{i}  \left( \epsilon \left( \alpha \left( g \right) \right) \right) \}_{i=1}^{r}$ is a basis of $A \overline{\Gamma}_{\epsilon \left( \alpha \left( g \right) \right)}^{T}$, we have that $\varphi^{X}$ is an immersion. Also, it satisfies that
$$\varphi^{X} \left( W \times U \right) \subseteq \overline{\Gamma}.$$
So, these charts give us an atlas over $\overline{\Gamma} \left( \mathcal{F} \left( x \right) \right)$ which induces a Hausdorff second contable topology on $\overline{\Gamma} \left( \mathcal{F} \left( x \right) \right) $ such that $\overline{\Gamma} \left( \mathcal{F} \left( x \right) \right)$ is an immersed submanifold of $\Gamma$. To end the proof we just have to use Eq. (\ref{23}) to prove that the source and the target mappings are submersions.\\
\hfill
\end{proof}
\end{theorem}

\begin{remark}
\rm
Our construction of the characteristic distribution associated to a subgroupoid $\overline{\Gamma}$ of a Lie groupoid $\Gamma$ can be seen as a generalization of the Lie functor which associates to any Lie groupoid a Lie algebroid (see \cite{KMG}). In fact, if $\overline{\Gamma}$ is a Lie subgroupoid of $\Gamma$, the chacteristic distribution $A \overline{\Gamma}^{T}$ induces the associated Lie algebroid to $\overline{\Gamma}$.\\
\hfill

\end{remark}

\section{Material groupoid and Material Distribution}

In this section we will apply the results of the section 3 to the case of continuum mechanics.\\
A \textit{body} $\mathcal{B}$ is a $3$-dimensional differentiable manifold which can be covered with just one chart. An embedding $\phi : \mathcal{B} \rightarrow \mathbb{R}^{3}$ is called a \textit{configuration of} $\mathcal{B}$ and its $1-$jet $j_{X,\phi \left(X\right)}^{1} \phi$ at $X \in \mathcal{B}$ is called an \textit{infinitesimal configuration at $X$}. We usually identify the body with any one of its configurations, say $\phi_{0}$, called \textit{reference configuration}. Given any arbitrary configuration $\phi$, the change of configurations $\kappa = \phi \circ \phi_{0}^{-1}$ is called a \textit{deformation}, and its $1-$jet $j_{\phi_{0}\left(X\right) , \phi \left(X\right)}^{1} \kappa$ is called an \textit{infinitesimal deformation at $\phi_{0}\left(X\right)$}.\\

For simple elastic bodies, the mechanical response of a material is completely characterized by one  function $W$ which depends, at each point $X \in \mathcal{B}$, on the gradient of the deformations evaluated at the point. Thus, $W$ is defined as a differentiable map
$$ W : \Pi^{1} \left( \mathcal{B}, \mathcal{B}\right) \rightarrow V,$$
which does not depend on the final point, i.e., for all $X,Y,Z \in \mathcal{B}$
\begin{equation}\label{11}
 W \left( j_{X,Y}^{1} \phi\right) = W \left( j_{X,Z}^{1} \left( \tau_{Z-Y} \circ \phi\right)\right), \ \forall j_{X,Y}^{1}\phi \in \Pi^{1} \left( \mathcal{B}, \mathcal{B}\right),
\end{equation}
where $V$ is a real vector space and $\tau_{v}$ is the translation map on $\mathbb{R}^{3}$ by the vector $v$. This map will be called \textit{response functional}. There are other equivalent definitions (for the other definitions see \cite{MELZA}, \cite{MELZASEG}, \cite{MEPMDLSEG} or \cite{VMJIMM}) of this function but we will use this one for convenience.\\


Now, just imagine that an infinitesimal neighbourhood of the material around the point $Y$ can be grafted so perfecly into a neighbourhood of $X$, that the graft cannot be detected by any mechanical experiment. If this condition is satisfied with every point $X$ of $ \mathcal{B}$, the body is said \textit{uniform}. We can express this physical property in a geometric way as follows.

\begin{definition}
\rm
A body $\mathcal{B}$ is said to be \textit{uniform} if for each two points $X,Y \in \mathcal{B}$ there exists a local diffeomorphism $\psi$ from an open neighbourhood $U \subseteq \mathcal{B}$ of $X$ to an open neighbourhood $V \subseteq \mathcal{B}$ of $Y$ such that $\psi \left(X\right) =Y$ and
\begin{equation}\label{12}
W \left( j^{1}_{Y, \kappa \left(Y\right)} \kappa \cdot j^{1}_{X,Y} \psi \right) = W \left( j^{1}_{Y, \kappa \left(Y\right)} \kappa\right),
\end{equation}
for all infinitesimal deformation $j^{1}_{Y , \kappa \left(Y\right)} \kappa$. $j^{1}_{X,Y} \psi$ is called a \textit{material isomorphism}.
\end{definition}
These kind of maps are going to be important and we will endow these maps of a groupoid structure over $\mathcal{B}$. For each two points $X,Y  \in \mathcal{B}$, we will denote by $G \left(X,Y\right)$ the collection of all $1-$jets $j_{X,Y}^{1}\psi$ which satisfy Eq. (\ref{12}). So, the set $\Omega \left( \mathcal{B}\right) = \cup_{X,Y \in \mathcal{B}} G\left(X,Y\right)$ can be considered as a groupoid over $\mathcal{B}$ which is, indeed, a subgroupoid of the $1-$jets groupoid $\Pi^{1}\left( \mathcal{B} , \mathcal{B}\right)$.\\
\indent{A \textit{material symmetry} at a point $X$ is a material isomorphism which takes $X$ to $X$. We denote by $G\left(X\right)$ the set of all material symmetries which is, indeed, the isotropy group of $\Omega \left( \mathcal{B} \right)$ at $X$. For each $X\in \mathcal{B}$, we will denote the set of material isomorphisms from $X$ to any other point (resp. from any point to $X$) by $\Omega_{X}\left( \mathcal{B}\right)$ (resp. $\Omega^{X}\left( \mathcal{B}\right)$). Finally, we will denote the structure maps of $\Omega \left( \mathcal{B} \right)$ by $\overline{\alpha}$, $\overline{\beta}$, $\overline{\epsilon}$ and $\overline{i}$ which are just the restrictions of the corresponding ones on $\Pi^{1} \left(  \mathcal{B} , \mathcal{B} \right)$.}\\
As a consequence of the continuity of $W$ we have that, for all $X \in \mathcal{B}$, $G \left( X \right)$ is a closed subgroup of $\Pi^{1} \left(\mathcal{B} , \mathcal{B}\right)_{X}^{X}$. Hence, the following result is immediate.
\begin{proposition}\label{28}
Let $\mathcal{B}$ be a simple body. Then, for all $X \in \mathcal{B}$ the set of all material symmetries $G \left( X \right)$ is a Lie subgroup of $\Pi^{1} \left( \mathcal{B} , \mathcal{B}\right)_{X}^{X}$.
\end{proposition}
This could make us think that $\Omega \left( \mathcal{B} \right)$ is a Lie subgroupoid of $\Pi^{1} \left( \mathcal{B} , \mathcal{B}\right)$. However, this is not true because the dimensions of the groups of material symmetries could change (see the examples of the last section).\\
Now, the following result is obvious.
\begin{proposition}
Let $\mathcal{B}$ be a body. $\mathcal{B}$ is uniform if and only if $\Omega \left( \mathcal{B}\right)$ is a transitive subgroupoid of $\Pi^{1} \left( \mathcal{B} , \mathcal{B}\right)$.
\end{proposition}
We will still deal with another (more restrictive) notion of uniformity.
\begin{definition}
\rm
A body $\mathcal{B}$ is said to be \textit{smoothly uniform} if for each point $X \in \mathcal{B}$ there is an neighbourhood $U $ around $X$ such that for all $Y \in U$ and $j_{Y,X}^{1} \phi \in \Omega \left( \mathcal{B} \right)$ there exists a smooth field of material isomorphisms $P$ from $\epsilon \left( X \right)$ to $j_{Y,X}^{1} \phi$.
\end{definition}
Equivalently, for all $ X \in \mathcal{B}$ the map
$$ \overline{\alpha} :\Omega^{X}\left( \mathcal{B}\right) \rightarrow \mathcal{B},$$
admits local sections for any point in $\mathcal{B}$. Obviously, smooth uniformity implies uniformity.\\
Therefore, $\mathcal{B}$ is smoothly uniform if, and only if, for each two points $X,Y \in \mathcal{B}$ there are two open subsets $U , V \subseteq \mathcal{B}$ around $X$ and $Y$ respectively and $P : U \times V \rightarrow \Omega \left( \mathcal{B} \right) \subseteq \Pi^{1} \left( \mathcal{B} , \mathcal{B} \right)$, a differentiable section of the anchor map $\left( \overline{\alpha} , \overline{\beta} \right)$. When $X=Y$ it is easy to realize that we can assume $U=V$ and $P$ is a morphism of groupoids over the identity map, i.e.,

$$ P \left( Z, T \right) = P \left(  R , T\right) P \left( Z, R \right), \ \forall T,R,Z \in U.$$ 

So, we may prove a corollary of Proposition \ref{28}.

\begin{corollary}
Let $\mathcal{B}$ be a body. $\mathcal{B}$ is smoothly uniform if and only if $\Omega \left( \mathcal{B}\right)$ is a transitive Lie subgroupoid of $\Pi^{1} \left( \mathcal{B} , \mathcal{B}\right)$.
\begin{proof}
Suppose that $\mathcal{B}$ is smoothly uniform. Fix $j_{X,Y}^{1} \psi \in \Omega \left( \mathcal{B} \right)$ and consider $P : U \times V \rightarrow \Omega \left( \mathcal{B} \right)$, a differentiable section of the anchor map $\left( \overline{\alpha} , \overline{\beta} \right)$ with $X \in U$ and $Y \in V$. Then, we may construct the following bijection

$$
\begin{array}{rccl}
\Psi_{U,V} : & \Omega \left( U,V\right) & \rightarrow & \mathcal{B} \times \mathcal{B} \times G \left( X , Y \right)\\
&j_{Z,T}^{1} \phi &\mapsto &  \left( Z,T , P \left( Z,Y \right) \left[ j_{Z,T}^{1} \phi \right]^{-1} P \left( X,T \right)\right)
\end{array}
,$$
where $\Omega \left( U,V\right)$ is the set of material isomorphisms from $U$ to $V$. By using Proposition \ref{28}, we deduce that $G \left( X , Y \right)$ is a differentiable manifold. Thus, we can endow $\Omega \left( \mathcal{B} \right)$ with a differentiable structure of a manifold. Now, the result follows (the converse has been proved in \cite{KMG}).

\end{proof}
\end{corollary}


\begin{remark}
\rm

Let $P$ be a left-invariant vector field on $\Pi^{1} \left( \mathcal{B} , \mathcal{B} \right)$ such that
\begin{equation}\label{13}
TW \left( P \right) = 0
\end{equation}
Then, for all $g \in \Pi^{1} \left( \mathcal{B} , \mathcal{B} \right)$, the flow $\varphi^{P}_{t}$ of $P$ satisfies that
\begin{eqnarray*}
TW \left( P \left( g \right)\right) &=& \dfrac{\partial}{\partial t_{|0}}\left(W \left(\varphi^{P}_{t}\left( g \right) \right)\right)\\
&=& \dfrac{\partial}{\partial t_{|0}}\left(W \left(g \cdot\varphi^{P}_{t}\left( \epsilon \left( \alpha \left( g \right) \right) \right) \right)\right) = 0.
\end{eqnarray*}

Hence, $\varphi^{P}_{t}\left( \epsilon \left( X \right) \right) \in \Omega \left( \mathcal{B} \right)$ for all $X \in \mathcal{B}$, i.e., the flow of $P$ restricts to $\Omega \left(  \mathcal{B}\right)$.\\
Equivalently, if the flow of $P$ can be restricted to $\Omega \left( \mathcal{B}\right)$ then, Eq. (\ref{13}) is satisfied.\\
Thus, let $g \in \Pi^{1} \left( \mathcal{B} , \mathcal{B} \right)$; then the fibre of the characteristic distribution $A \Omega \left( \mathcal{B} \right)^{T}_{g}$ to the material groupoid at $g$ is generated by the vectors $v_{g} \in T_{g} \beta^{-1} \left( \beta \left( g \right)\right)$ such that there exists a left-invariant (local) tangent vector field $ P \in \frak X_{L} \left( \Pi^{1} \left( \mathcal{B} , \mathcal{B} \right) \right)$ which satisfies Eq. (\ref{13}) and $  P \left( g\right) = v_{g}$. So, for the material groupoid, the characteristic distribution is defined in a more straightforward way as the distribution generated by the left-invariant vector fields on $\Pi^{1} \left( \mathcal{B} , \mathcal{B} \right)$ which are in the kernel of $TW$.\\
This characteristic distribution will be called \textit{material distribution}. The distribution on the base $A \Omega \left( \mathcal{B} \right)^{\sharp}$ will be called \textit{body-material distribution}. Denote again the family of left-invariant vector fields on $\Pi^{1} \left( \mathcal{B} , \mathcal{B} \right)$ which satisfy Eq. (\ref{13}) by $\mathcal{C}$.\\

\hfill
\end{remark}
Let $\overline{\mathcal{F}} \left( \epsilon \left( X \right) \right)$ and $\mathcal{F} \left( X \right)$ be the foliations associated to the material distribution and the body-material distribution respectively.  For each $X \in \mathcal{B}$, we denote the Lie groupoid $\Omega \left( \mathcal{B} \right)\left(\mathcal{F}\left( X \right)\right)$ by $\Omega \left( \mathcal{F} \left( X \right) \right)$.\\
\indent{Notice that, strictly speaking, in continuum mechanics a \textit{subbody} of a body $\mathcal{B}$ is just an open submanifold of $\mathcal{B}$ but, here, the foliation $\mathcal{F}$ gives us submanifolds of different dimensions. So, we will consider a more general definition so that, a \textit{material submanifold of $\mathcal{B}$} is just a submanifold of $\mathcal{B}$. A generalized subbody $\mathcal{P}$ inherits certain material structure from $\mathcal{B}$. In fact, we will measure the material response of a material submanifold $\mathcal{P}$ by restricting $W$ to the $1-$jets of local diffeomorphisms $\phi$ on $\mathcal{B}$ from $\mathcal{P}$ to $\mathcal{P}$. However, it easy to observe that a material submanifold of a body is not exactly a body. See \cite{MD} for a discussion on this subject.}

Then, as a corollary of Theorem \ref{20}, we have the following result.
\begin{theorem}
For all $X \in \mathcal{B}$, $\Omega \left( \mathcal{F} \left( X \right) \right)$ is a transitive Lie subgroupoid of $\Pi^{1} \left( \mathcal{B} , \mathcal{B} \right)$. Thus, any body $\mathcal{B}$ can be covered by a foliation of smoothly uniform material submanifolds.
\end{theorem}
Notice that, for each $X \in \mathcal{B}$, all the material symmetries at $X$ are contained in $\Omega \left( \mathcal{F} \left( X \right) \right)$, i.e., $G \left( X \right)$ is the isotropy group at $X$ of $\Omega \left( \mathcal{F} \left( X \right) \right)$.}\\

\begin{remark}
\rm
Just imagine that there is, at least, a leaf $\overline{\mathcal{F} }\left( g \right)$ inside $\Omega^{X} \left( \mathcal{B} \right)$ for some $X \in \mathcal{B}$ such that
$$ \overline{\mathcal{F}} \left( g \right) \cap \overline{\mathcal{F}} \left( \epsilon \left( X \right) \right)= \emptyset.$$
Then, we are not including this leaf inside any subgroupoid $\Omega \left( \mathcal{F} \left( X \right) \right)$. Thus, we are discarding these material isomorphisms.\\
However,
\begin{equation}\label{16}
\overline{\mathcal{F}} \left( g \right) = g \cdot \overline{\mathcal{F}}\left( \epsilon \left( \alpha \left( g \right) \right) \right),
\end{equation}
and $ \overline{\mathcal{F}}\left( \epsilon \left( \alpha \left( g \right) \right) \right)$ is included in $\Omega \left( \mathcal{F} \left( \alpha \left( g \right) \right)\right)$, i.e., using Eq. (\ref{16}), we can reconstruct $\overline{\mathcal{F}} \left( g \right)$.
\hfill
\end{remark}


\section{Examples}
Finally, let us introduce a large family of examples. We will see that, in some of them, the material groupoid is not a Lie groupoid (this kind of examples justify the study of groupoids without structure of Lie groupoids). We shall also give the decomposition of the material by smoothly uniform material submanifolds provided by the characteristic distribution. In these examples we will also show that the leaves $\overline{\mathcal{F}} \left( \epsilon \left( X \right) \right)$ are contained in the $\overline{\beta}-$fibres of $\Omega \left( \mathcal{B} \right)$ but they do not coincide in general.\\
Let $\mathcal{B}$ be an open ball $B_{r} \left( 0 \right)$ in $\mathbb{R}^{3}$ of radius $r$ and center $0 \in \mathbb{R}^{3}$ whose functional response is given by
$$
\begin{array}{rccl}
W : & \mathcal{B} \times Gl \left( 3 , \mathbb{R} \right) & \rightarrow & \frak gl \left( 3 , \mathbb{R} \right) \\
& \left(X,Y,Z, F \right) &\mapsto &  f \left( \parallel \left(X,Y,Z  \right) \parallel^{2}\right) \left(  F\cdot F^{T} - Id\right)
\end{array}
,$$
where $f: \mathbb{R} \rightarrow \mathbb{R}$ is a differentiable map. Here, we will consider different cases:
\begin{itemize}
\item[\textbf{(1)}] \textbf{$f$ is strictly monotonic.}\\
Then, two points $\left( X , Y , Z \right), \left( T,S,R \right)$ of $\mathcal{B}$ are materially isomorphic if, and only if, 
$$ \parallel \left(X,Y,Z  \right) \parallel = \parallel \left(T,S,R  \right) \parallel.$$
In fact, we have
$$ \Omega \left( \mathcal{B} \right)^{\left(X,Y,Z \right)} \cong S \left( \parallel \left(X,Y,Z \right) \parallel \right) \times O \left( 3 \right),$$
where $S \left( \parallel \left(X,Y,Z \right) \parallel \right)$ is the sphere of radius $\parallel \left(X,Y,Z \right) \parallel$ and centre $0$ and $O \left( 3 \right)$ is the orthogonal group of $3 \times 3$ matrices with coefficients in $\mathbb{R}$.\\
Hence, considering local coordinates over the sphere we can prove that $\Omega \left( \mathcal{B} \right)$ is, indeed, an embedded Lie subgroupoid of $\Pi^{1} \left( \mathcal{B} , \mathcal{B} \right)$. The material distribution (which, in this case, is a Lie algebroid) is given by
$$ A \Omega \left( \mathcal{B} \right)^{T}_{\epsilon \left( X,Y,Z\right) } \cong \mathbb{R}^{2} \times \frak o \left( 3 \right),$$
for all $\left( X,Y,Z\right) \in \mathcal{B}$ where $\frak o \left( 3 \right)$ is the Lie algebra associated to the orthogonal group $O \left( 3 \right)$. Thus, the decomposition in material submanifolds is given by spheres of radius $< r$. Here, it is easy to construct non-trivial examples of left-invariant vector fields tangent to the characteristic distribution. In fact, any differentiable map $ \mathbb{R}^{3}  \rightarrow  \mathbb{R}^{2} \times \frak o \left( 3 \right)$ generates (local) left-invariant vector fields tangent to $A \Omega \left( \mathcal{B} \right)^{T}$.

\item[\textbf{(2)}] \textbf{$f$ is strictly monotonic and locally constant.}\\
In this case, we will suppose that there exists $s<r$ such that
$$ {f}_{| \left[0, s^{2}\right]} \equiv 1,$$
and is strictly monotonic non-zero outside of $\left[ 0,s^{2} \right]$. Denoting the closure of $B_{s} \left( 0 \right)$ by $\overline{B}_{s} \left( 0 \right)$, we have that

\begin{itemize}
\item For $\left(X,Y,Z \right) \in \overline{B}_{s}\left( 0 \right)$,
$$ \Omega \left( \mathcal{B} \right)^{\left(X,Y,Z \right)} \cong \overline{B}_{s}\left( 0 \right) \times O \left( 3  \right).$$
\item For $\left(X,Y,Z \right) \notin \overline{B}_{s}\left( 0 \right) $.
$$ \Omega \left( \mathcal{B} \right)^{\left(X,Y,Z \right)} \cong S \left( \parallel \left(X,Y,Z \right) \parallel \right) \times O \left( 3 \right).$$
\end{itemize}
Thus, the material groupoid is the following

\begin{small}
\begin{equation}\label{27}
\Omega \left( \mathcal{B} \right) \cong \{ \left( X,Y,h \right) \ : \ h \in \mathcal{O} \left(3 \right), \ \parallel X \parallel = \parallel Y \parallel \ \text{or} \  \parallel X \parallel ,\parallel Y \parallel \leq s \},
\end{equation}
\end{small}

\noindent{therefore, the dimension of the $\overline{\beta}-$fibres of the material groupoid is not constant and, then, the material groupoid cannot be a Lie groupoid. In fact, as we can see in Eq. \ref{27}, the material groupoid is not a submanifold of $\Pi^{1} \left( \mathcal{B} , \mathcal{B} \right)$.}\\
\indent{As we predicted, this example also shows the leaves $\overline{ \mathcal{F}} \left( \epsilon \left( X \right) \right)$ could be different from the $\overline{\beta}-$fibres of $\Omega \left( \mathcal{B} \right)$ (because in this case the $\overline{\beta}-$fibres are not manifolds without boundary).}\\
Now, the material distribution satisfies
\begin{itemize}
\item[(i)] $ A \Omega \left( \mathcal{B} \right)^{T}_{\epsilon \left( X,Y,Z\right) } \cong \mathbb{R}^{3} \times \frak o \left( 3 \right),$\\
for all $\left( X,Y,Z\right) \in B_{s}\left( 0 \right)$.
\item[(ii)] $ A \Omega \left( \mathcal{B} \right)^{T}_{\epsilon \left( X,Y,Z\right) } \cong  \mathbb{R}^{2} \times \frak o \left( 3 \right),$\\
for all $\left( X,Y,Z\right) \in  \left( \mathcal{B} - B_{s}\left( 0 \right) \right) $.
\end{itemize}
Notice that, because of the left-invariance, the material distribution is characterized by the fibres at the identities.\\
With this, the decomposition of $\mathcal{B}$ given by uniform material submanifolds is the following:
$$\mathcal{F} = \{  B_{s}\left( 0 \right), S \left( \parallel T \parallel \right) \ : \ T \in  \left( \mathcal{B} - B_{s}\left( 0 \right) \right) \}.$$
To prove this fact it is not difficult to check the following:
\begin{itemize}
\item[-] Let be a point $\left( X,Y,Z\right) \in B_{s}\left( 0 \right)$. So, by taking an open neighbourhood $U$ of $\left( X,Y,Z \right)$ contained in $B_{s}\left( 0 \right)$, we prove $(i)$.\\
\item[-] Any (local) vector field $P$ tangent to the characteristic distribution satisfy that its flow $\varphi^{P}_{t}$ at a point $\left( X,Y,Z\right)$ outside $B_{s}\left( 0 \right)$ is contained in $ S \left( \parallel \left( X,Y,Z\right) \parallel \right)$. This proves $(ii)$.
\end{itemize}

\item[\textbf{(3)}] \textbf{$f$ is neither monotonic nor locally constant.}\\
In this case, two points $\left( X , Y , Z \right), \left( T,S,R \right)$ of $\mathcal{B}$ are materially isomorphic if, and only if, 
$$ \parallel \left(X,Y,Z  \right) \parallel^{2} \in f^{-1} \left( f \left( \parallel \left(T,S,R  \right) \parallel^{2} \right) \right).$$
So, for each $\left( R,T,S\right) \in \mathcal{B}$, we have
$$ \Omega \left( \mathcal{B} \right)^{\left( T,S,R \right)} \cong \sqcup_{\parallel \left(X,Y,Z  \right) \parallel^{2} \in f^{-1} \left( f \left( \parallel \left(T,S,R  \right) \parallel^{2} \right) \right)} S \left( \parallel \left(X,Y,Z \right) \parallel \right) \times O \left( 3 \right).$$

Fix a point $\left( X , Y, Z \right) \in \mathcal{B}$ such that
$$\dfrac{\partial \ f}{\partial t_{|\parallel \left(X,Y,Z \right) \parallel^{2}}} \lessgtr 0.$$ 
Then, $f$ is not constant and, hence, this kind of points have to exist.\\
So, by using the inverse function theorem, there is an open neighbourhood $U$ around $\parallel \left(X,Y,Z \right) \parallel^{2}$ such that $f_{|U}$ is a diffeomorphism onto its image. Then, the intersection of $U \times U \times Gl \left( 3 , \mathbb{R} \right)$ (which is an open set of $\Pi^{1} \left( \mathcal{B} , \mathcal{B} \right)$) with the material groupoid is given by
$$ \Omega \left( U \right) := \{ \left(\left( X,Y,Z \right) , \left( T, S, R \right), F^i_j \right) \}_{F^i_j \in \mathcal{O} \left( 3 \right), \ X,Y,Z,T,S,R \in U, \  \parallel \left(X,Y,Z \right) \parallel= \parallel \left(T,S,R  \right) \parallel }.$$
Hence, if it there exists a differentiable structure over $\Omega \left( \mathcal{B} \right)$ such that it is a Lie subgroupoid of $\Pi^{1} \left( \mathcal{B} , \mathcal{B} \right)$, $\Omega \left( U \right)$ will be an open subset and, in this way, $\Omega \left( \mathcal{B}\right)$ will have dimension $8$.\\
So, we can ensure that, $\Omega \left( \mathcal{B} \right) $ is a Lie subgroupoid of $\Pi^{1} \left( \mathcal{B} , \mathcal{B} \right)$ if, and only if, for all $\left( T,S,R  \right) \in \mathcal{B}$ $f^{-1} \left( f \left( \parallel \left(T,S,R  \right) \parallel^{2} \right) \right)$ has a countable number of connected components (by using the fact that manifolds are second countable).\\
Notice that, in this case, we can find a map $f$ such that the decomposition is still given by spheres (all with the same dimension) but the material groupoid does not have a differentiable structure.
\end{itemize}
\vspace{0.1cm}

\bibliographystyle{plain}
\bibliography{Library}

\end{document}